\documentclass[aps,pra,10pt,letterpaper,superscriptaddress,twocolumn,nofootinbib]{revtex4-1}

\usepackage{amsmath,amssymb,amsfonts,times,xcolor,graphicx}
\usepackage[uselistings]{revquantum}
\usepackage[bf]{subfigure}

\definecolor{darkblue}{RGB}{0,0,127} % choose dark colors for high contrast
\definecolor{darkgreen}{RGB}{0,150,0}
\hypersetup{
	breaklinks, 
	colorlinks, 
	linkcolor=darkblue, 
	citecolor=darkgreen, 
	filecolor=red, 
	urlcolor=blue,
	pdftitle={Tailored codes for small quantum memories}, 
	pdfauthor={A. Robertson, C. Granade, S. D. Bartlett, and S. T. Flammia}
}

\newcommand{\figurefolder}{./fig}

\begin{document}
\def\equationautorefname~#1\null{Eq.\ (#1)\null}
\def\figureautorefname{Fig.\@}
\def\subfigureautorefname{Fig.\@}
\subfigcapskip = 2pt
\subfigbottomskip = -5pt

\title{Tailored codes for small quantum memories}

\author{Alan Robertson}
\author{Christopher Granade}
\author{Stephen D. Bartlett}
\affiliation{Centre for Engineered Quantum Systems, School of Physics, The University of Sydney, Sydney, Australia}
\author{Steven T. Flammia}
\affiliation{Centre for Engineered Quantum Systems, School of Physics, The University of Sydney, Sydney, Australia}
\affiliation{Center for Theoretical Physics, Massachusetts Institute of Technology, Cambridge, USA}

\date{19 September 2017}

\begin{abstract}
	We demonstrate that small quantum memories, realized via quantum error correction in multi-qubit devices, can benefit substantially by choosing a quantum code that is tailored to the relevant error model of the system. 
	For a biased noise model, with independent bit and phase flips occurring at different rates, we show that a single code greatly outperforms the well-studied Steane code across the full range of parameters of the noise model, including for unbiased noise.  
	In fact, this tailored code performs almost optimally when compared with 10,000 randomly selected stabilizer codes of comparable experimental complexity. 
	Tailored codes can even outperform the Steane code with realistic experimental noise, and without any increase in the experimental complexity, as we demonstrate by comparison in the observed error model in a recent 7-qubit trapped ion experiment. 
\end{abstract}

\maketitle

%----------------------------------------------------------------------------------------------------------------------

\section{Introduction}

Complete control over small-scale quantum devices has now been demonstrated in a number of physical systems. 
Both trapped ions and superconducting qubits have shown single- and two-qubit gate operations with high fidelity on systems with fewer than $10$ qubits~\cite{Nigg_2014, kelly_state_2015, corcoles_demonstration_2015, veldhorst_two-qubit_2015, debnath_demonstration_2016}; many other architectures show promise for this capability in the near future. 
With such systems, the basic operations of quantum error correction have been demonstrated on quantum codes consisting of 3 to 7 qubits~\cite{reed_realization_2012, kelly_state_2015, corcoles_demonstration_2015, moussa_demonstration_2011, cory_experimental_1998, linke_2016}, including the 7-qubit Steane code that is often studied in quantum computing architectures~\cite{Nigg_2014}. 

A central goal of such experiments is to demonstrate that quantum error correction can be of net benefit, i.e., it can yield a logical error rate that is lower than the bare physical error rate. 
Achieving this goal is a challenge, in part because the commonly used quantum codes are not always well suited for the task. 
Many quantum codes---large and small---as well as their decoders are optimized for performance under noise models that are rarely exhibited in practice. 
The standard noise model for studying quantum error correcting codes and fault-tolerant circuits is for stochastic single-qubit Pauli $X$, $Y$, and $Z$ errors to occur with equal probability, i.e., a uniform single-qubit depolarization channel, independently on all qubits. 
While this noise model has a theoretical simplicity, it is not representative of what is seen in experiments. 
For example, in trapped ions~\cite{Nigg_2014}, quantum dots~\cite{Shulman_2012}, certain variants of superconducting qubits~\cite{Aliferis_2009}, and other qubits defined by nondegenerate energy levels with a Hamiltonian proportional to $Z$, the noise model is generically described by a dephasing ($Z$-error) rate that is much greater than the rates for relaxation and other non-energy-preserving error.
Correlated errors and non-Markovian noise are also pervasive in real devices~\cite{Epstein, Fogarty, Ball2015}. 

These experiments raise several new questions in the theory of quantum error correction.
In particular, as many of these experimental demonstrations show evidence of error models beyond the simplistic paradigm generally used to compare error-correcting codes, we may ask if it is possible to identify better choices of codes for these more general error models.
Moreover, it is important to assess whether a given error-correcting code and decoder are specific to a particular error model, or if they work well across a range of physical models.
Finally, we must be able to intelligently trade off between error-correcting performance for realistic error models and other experimental design concerns, such as minimizing the complexity of performing syndrome measurements (quantified, say, by the weight of stabilizer group generators).

In this work, we address all of these concerns and show that several current and near-future experiments will benefit significantly by tailoring small codes and their decoders for a more realistic noise model. 
We focus on two noise models: the first is a simple uncorrelated noise model with bias, and the second is based on the ion trap qubit experiments of \citet{Nigg_2014}, where the noise is both significantly biased towards dephasing and exhibits correlations.
We note that our results are widely applicable, as the \citet{Nigg_2014} error model serves as a representative example for similar reasoning in other experiments.
We study a class of 7-qubit stabilizer codes that require identical resources to the commonly used Steane code: all of the syndrome measurements require measuring at most four-body operators. 
We find that a simple tailored code outperforms the Steane code across \emph{the entire range} of bias parameters when used with the optimal decoder. 
We perform a case study of this tailored code for the trapped ion qubit experiment of \citet{Nigg_2014} and compare the performance of the tailored code, the 7-qubit Steane code, and the best stabilizer code selected from a set of 10,000 randomly generated codes (subject to the weight constraint above). 
We show that the tailored code can substantially outperform the Steane code under the observed error model of this experiment, under certain assumptions about the frequency of multi-qubit errors.
Furthermore, we show that there is still scope for substantial additional improvement using the random codes, despite them having no additional complexity in their experimental implementation. 

%----------------------------------------------------------------------------------------------------------------------

\section{Decoders} 

Quantum information can be protected from errors by encoding into a larger system, and using carefully-chosen measurements to identify when and where errors have occurred~\cite{nielsen00}.  
We restrict our consideration to \emph{stabilizer codes}~\cite{GottesmanPhD}, where the logical subspace is defined as the joint $+1$ eigenspace of a set of commuting Pauli operators called the \emph{stabilizers}.
If an error occurs, it will typically not commute with one or more of these stabilizers and so will flip their values.  
Measuring the stabilizers yields a set of \emph{error syndromes}, which is a list of those stabilizers that return the value $-1$.  
Moreover, this syndrome measurement projects the noise into an eigenstate of the stabilizer operators, thus discretizing the noise and allowing for digital error correction. 

A \emph{decoder} then attempts to identify the error by using the information inferred from the set of error syndromes.  Specifically, a decoder functions by identifying the most probable error given a set of observed syndromes. 
This identification clearly depends on the \emph{a priori} information available to the decoder regarding the underlying error model. 
Considering different error models, then, can lead to different decoders even for the same quantum code.

For a given quantum code, the noise model determines the rate for various errors, some of which are correctable and some are not. 
Each error results in a set of syndromes, and the decoder uses this syndrome information to determine the most likely error and correction. 
With this information, we can calculate the rate at which an optimal decoder fails to recover the correct logical state. 
A common notion of the optimal decoder is the \emph{maximum likelihood decoder}. 
It maximizes the \emph{a posteriori} probability of a given logical error conditioned on an observed syndrome, and this is the notion of optimal that we adopt. 
Throughout this paper, we perform a brute force numerical search for the optimal decoding operation over all possible recovery operations. 
As our focus is on small codes, this numerical approach is tractable.

Consider the case of a biased Pauli error model~\cite{Aliferis_2008, Aliferis_2009, Webster15, Tuckett_2017}, which will serve as our primary example. 
In this error model, independent Pauli $X$ and $Z$ errors occur at different rates, $r_x$ and $r_z$ respectively. 
The probability of an $X$ error is therefore $p_x = r_x (1-r_z)$ and for a $Z$ error, $p_z = r_z (1-r_x)$.
A $Y$ error occurs with probability $p_y = r_x r_z$. 
We define the total error probability $p$ to be the probability of any of $X, Y$, or $Z$ occurring, $p = p_x + p_y + p_z$. 
We assume $p_x \leq p_z$, that is, that $Z$ errors occur more often than $X$ errors, and we define the \emph{bias} to be $\eta = p_z/p_x$. 
Note that there is not a unique convention for defining bias in the literature. 
We follow the definition of \citet{Webster15} rather than \citet{Aliferis_2008} and \citet{Aliferis_2009} since we are not considering explicit fault-tolerant gates.  
For more complex noise models such as for coherent noise, the notion of error rate becomes more ambiguous, but it is well-defined for our stochastic noise model. 
The most useful parameterization of our error channel is by the total error rate $p$ together with the bias $\eta$ rather than the individual $X$ and $Z$ error probabilities $p_x$ and $p_z$ or the bare error rates $r_x$ and $r_z$. 
However, all of these parameterizations are equivalent. 
By logical error rate, we mean the probability that any logical error has occurred. 

A simple toy picture illustrates how adapting a code and decoder to a biased noise channel can offer lower logical error rates. 
Suppose two errors, $X$ and $ZZ$ have the same syndrome, but lie in different logical classes.
In an unbiased model ($\eta = 1$), $X$ is the more likely error since it has lower weight. 
In a sufficiently biased regime, however, $ZZ$ will become more likely. 
Supposing further that contributions from other errors are negligible, a code that can correct two-qubit $ZZ$ errors and a decoder that is adapted to this channel bias will outperform the original code and decoder that was used for the unbiased noise. 
What is perhaps surprising is that this approach can lead to significant improvements even for small codes, as our results demonstrate. 

%----------------------------------------------------------------------------------------------------------------------

\section{Tailored codes} 

Our goal is to tailor a code and decoder to a specific error model on a fixed number of qubits to minimize the logical error rate. 
As our focus is on small, physically realizable quantum memories, we need to impose some additional constraints. 
We want the syndrome to be easy to measure, so we restrict to stabilizer codes having generators of weight at most four. 
This facilitates a direct comparison to Steane's code~\cite{Steane1996}, defined by the stabilizer generators
\begin{equation}
\label{eq:Steane}
{\scriptsize
\begin{array}{ccccccc}
 	X & I & X & I & X & I & X, \\
	I & X & X & I & I & X & X, \\ 
	I & I & I & X & X & X & X, 
\end{array} }
\ + \ (X \to Z)\,.
\end{equation}
Furthermore, we require that our code can correct at least one arbitrary quantum error, so we restrict to codes with distance at least three. 
Since we focus on memories only, and not fault-tolerant quantum computation, we are not concerned with the number of transversal gates that the code can implement; it is here that our codes may be inferior to Steane's code. 

What is the best stabilizer code for the biased Pauli error model? 
It seems unlikely that there is a single optimal code for all choices of bias. 
However, we have found that a specific 7-qubit cyclic stabilizer code performs nearly optimally for any bias. 
The code is defined by the stabilizers
\begin{align}
\label{eq:Cyclic}
	XZIZXII, \ + \text{ cyclic}\,.
\end{align}
This cyclic code has connections to a range of other commonly used stabilizer codes. 
First, its stabilizers are a simple generalization of those of the perfect 5-qubit stabilizer code ($XZIZX +$ cyclic) to 7 qubits, by padding with identity operators. 
In addition, this 7-qubit code, like the 5-qubit code, are the smallest toric codes defined on a toroidal lattice with shifted periodic boundary conditions (specifically, they are equivalent to toric codes following the variation detailed by Wen~\cite{WenPlaquette}). 
Third, they relate to a well known quantum convolutional code~\cite{Ollivier2003}, but with periodic boundary conditions. 

\begin{figure*}[t!]
	\begin{center}
		\subfigure[]{
			\label{subfig:p}
			\includegraphics[width=0.95\columnwidth]{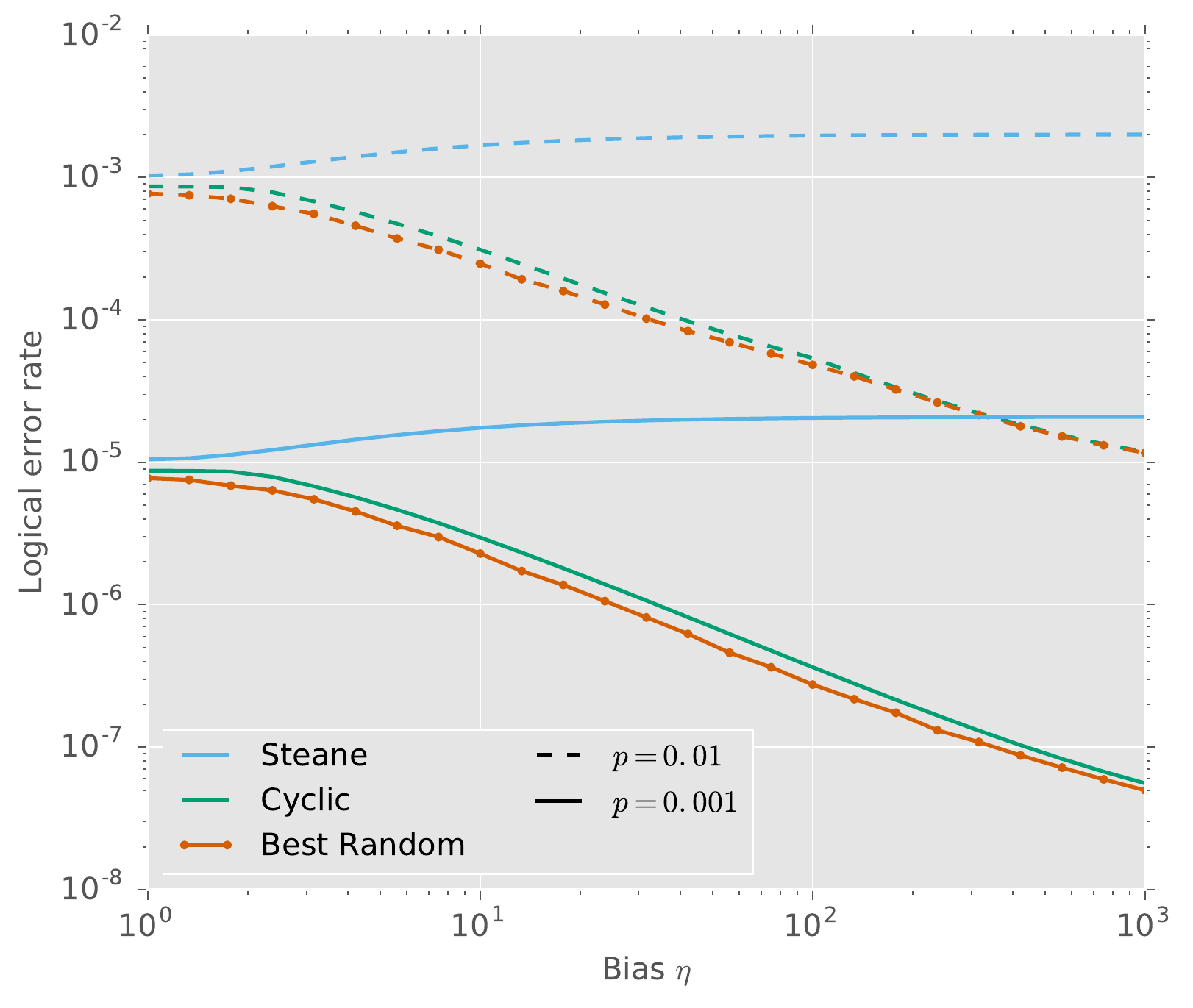}
		} \qquad
		\subfigure[]{
			\label{subfig:bias}
			\includegraphics[width=0.95\columnwidth]{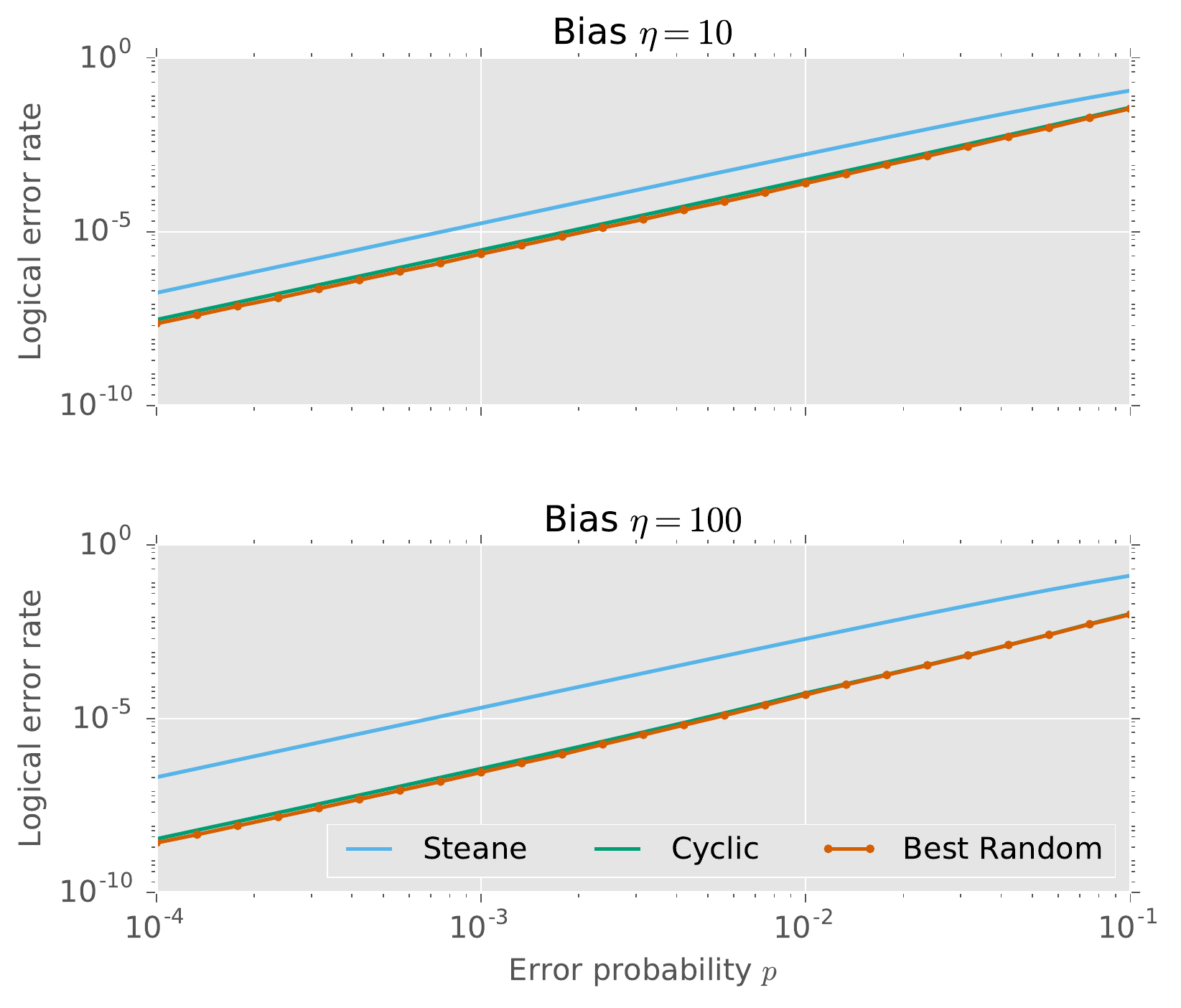}
		}
	\end{center}
	\caption{
		\label{fig:7_qubit_bias}
		\textbf{(a)} The probability of a logical error for different 7-qubit stabilizer codes under a biased Pauli error model as the bias is varied for two choices of physical error rate, $p = 0.01$ and $0.001$. 
		The blue line and green line denote the performance of the Steane code and the cyclic code of \autoref{eq:Cyclic} respectively. 
		The red show the performance of the best identified random code for that error rate. 
		\textbf{(b)} The probability of a logical error for different 7-qubit stabilizer codes under a biased Pauli error model as a function of $p$ for two values of the bias.
	}
\end{figure*}

Using a biased Pauli error model, we compare this 7-qubit cyclic code of \autoref{eq:Cyclic} to the Steane code (the smallest instance of a two-dimensional color code~\cite{Bombin_2006}).
We also compare against the best code using an optimal decoder among a randomly chosen sample of 10,000 stabilizer codes obeying our restrictions of having generators with weight at most four and distance at least three, referred to as the \emph{best random code}. 

Remarkably, our tailored 7-qubit code outperforms the Steane code against a biased Pauli error model for all values of bias and total physical error rate, as shown in \autoref{fig:7_qubit_bias}, including the case when there is no bias. 
While the best random code outperforms both, as required, we note that we have reidentified the best random code for each value of bias. 
We emphasize that the \emph{fixed} tailored code performs well over the entire range, and is indeed quite close in performance to the best random code for each parameter value. 
This performance is for a maximum likelihood decoder that depends on the error model; that is, the decoder varies as a function of bias. 
A specific choice of decoder that is independent of the channel parameters can still be optimal over a range of noise parameters, of course. 
Because there is some degree of insensitivity in the decoder behavior when the channel parameters are changed, we do not require precise knowledge of the channel parameters to achieve near-optimal performance. 
This can also be exploited by making use of adaptive decoders, as has been investigated in  \cite{combes_situ_2014,fowler_scalable_2014, florjanczyk_-situ_2016}, to learn the optimal choice of decoding function from observed syndrome measurements.

As noted above, the 7-qubit cyclic code of \autoref{eq:Cyclic} is a small instance of a variant of the toric code, and as such is part of a family of high-threshold codes up to arbitrary size.  A very recent result demonstrates that such a family of topological codes, tailored appropriately, can maintain its exceptional performance across all values of bias~\cite{Tuckett_2017}, consistent with the results shown here.

%----------------------------------------------------------------------------------------------------------------------

\section{Noise models for trapped ion qubits} 

\begin{figure*}[t!]
	\begin{center}
		\subfigure[]{
			\label{subfig:qpt-point-est}
			\includegraphics[width=0.8\columnwidth]{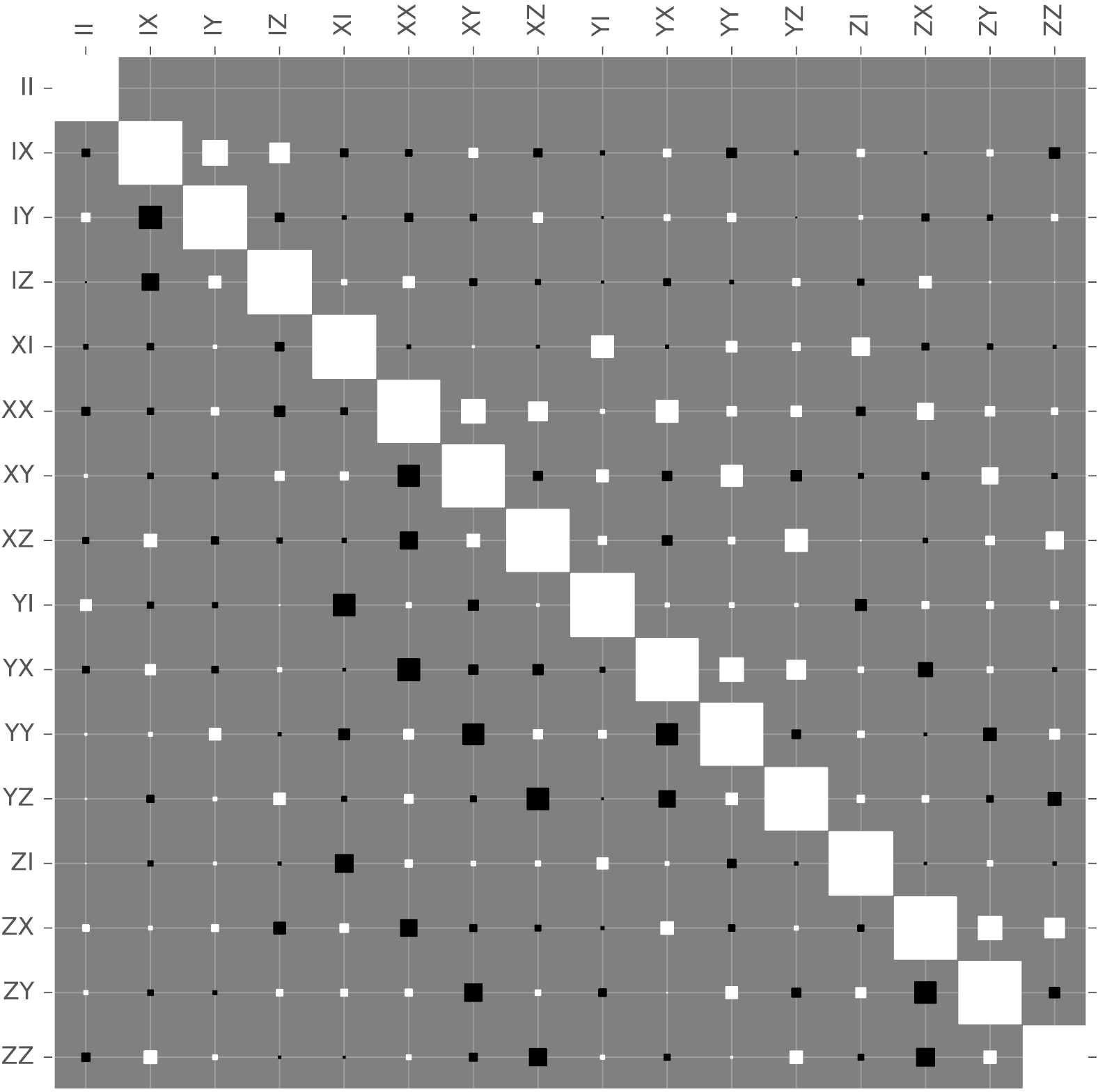}
		} \qquad
		\subfigure[]{
			\label{subfig:innsbruck}
			\includegraphics[width=0.98\columnwidth]{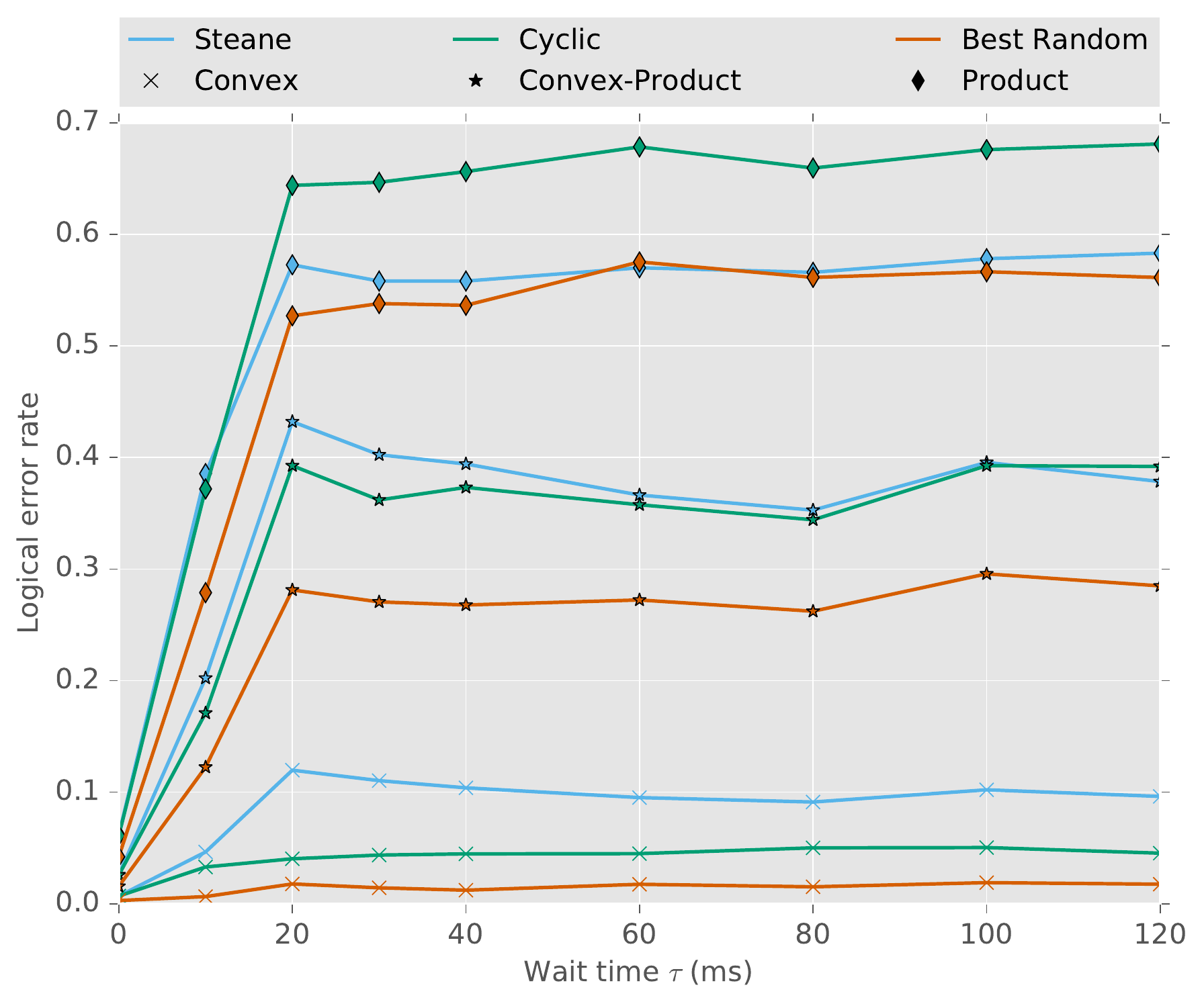}
		}
	\end{center}
	\caption{
		\label{fig:innsbruck}
		\textbf{(a)} The superoperator representation, in the Pauli basis, of the channel estimated from process tomography data collected from the Innsbruck experiment. 
		The tomography experiment consists of preparing and then immediately measuring a state (that is, a wait time of $\tau = 0$), as described in the supplementary material \cite{supplemental}. 
		This Hinton diagram depicts each positive (negative) matrix element by a white (black) square, where the size of each square indicates the relative magnitude of the respective matrix elements.
		The presence of non-negligible off-diagonal elements, such as $(XY,XX)$, indicates that the estimated channel cannot be entirely explained as a convex combination of the action of Pauli operators. 
		The first row and column indicate that the estimated channel is trace preserving, but not unital. 
		\textbf{(b)} The probability of a logical error following the application of a recovery operation for the Innsbruck tomography data compared to the time over which the state is left to decohere. 
		These data include a high rate of correlated two-qubit $Z$ errors. 
		This figure demonstrates that the use of a tailored stabilizer code with an optimal decoding that corrects for two qubit Z errors greatly exceeds the correction ability of an optimally decoded Steane code. 
	}
\end{figure*}

To demonstrate the utility of tailoring codes in practical situations, we now present a case study of tailored 7-qubit codes for a noise model obtained from the Innsbruck ion trap experiment~\cite{Nigg_2014}. 
We use the results of two-qubit process tomography of the ``identity channel'' (i.e., no gate operations) performed in this experiment for various waiting times ranging $\tau$ from $0$ to $120$ ms. 
From these data, we reconstruct the 2-qubit noisy error channel for each time. 
Our channel reconstruction for the shortest waiting time $\tau = 0$ is shown in \autoref{subfig:qpt-point-est}.

The two-qubit tomography data only give partial information about the noise model for a 7-qubit device, and we must extrapolate the error model in some way.
Given the tomographic estimate $\hat{\Lambda}$ of the error channel at each time, we first approximate the channel using the Pauli-twirl, $\mathcal{E} = \sum_{P \in \mathcal{P}} P \hat{\Lambda} P^\dagger$, where $\mathcal{P}$ is the two-qubit Pauli group.
Though in some important cases the Pauli-twirling approximation leads to underestimates of the total error rates \cite{puzzuoli_tractable_2014} and pseudothresholds \cite{gutierrez_comparison_2015}, we are only using the twirled channels to derive plausible noise distributions.
Let $\mathcal{E}_{ij}$ be the action of $\mathcal{E}$ on a pair of qubits $i,j$.
We consider three alternative extrapolations of $\mathcal{E}_{ij}$ to the full seven-qubit register:
\textit{(i)~Convex}. The noise model is a convex combination of the 2-qubit channel on a pair of the 7 qubits, with the identity on all others, sampled uniformly over all choice of neighboring pairs on a line, i.e., 
$\mathcal{E}_{c} = \tfrac{1}{6}\sum_{j=1}^{6} \mathcal{E}_{j,j+1}$. 
\textit{(ii)~Convex product.} The noise model is a convex combination of the 2-qubit channel acting independently on even or on odd pairs, i.e., 
$\mathcal{E}_{cp} = \frac{1}{2} \mathcal{E}_{12} \mathcal{E}_{34} \mathcal{E}_{56} + \frac{1}{2} \mathcal{E}_{23} \mathcal{E}_{45} \mathcal{E}_{67}$. 
\textit{(iii)~Product.} The noise model is a product of the 2-qubit channel on all neighboring pairs on a line, i.e., 
$\mathcal{E}_{p} = \prod_{j=1}^{6} \mathcal{E}_{j,j+1}$, where the product is unambiguous because the channels are Pauli channels and hence commute.
We note that the convex extrapolation likely underestimates the noise occurring in a 7-qubit experiment, and the product extrapolation likely overestimates it; the convex product extrapolation represents a level of noise that is in between these two extremes. 
We also note that these extrapolations differ substantially in the frequency of high-weight errors. 
The convex extrapolation contains only single- and two-qubit errors; the other two contain higher-weight errors and in particular the product extrapolation has a very substantial high-weight error rate. 
As our tailored code has been selected based only on consideration of the low-weight errors of the biased Pauli error model, we expect its relative performance to vary considerably depending on the choice of extrapolation.
The details of our tomographic reconstruction implementation and of our extrapolations onto the full seven-qubit register are provided in the supplemental material \cite{supplemental}~as a Jupyter Notebook~\cite{jupyter_development_team_jupyter_2016}.
We performed our reconstruction and extrapolation using QuTiP 4.1.0~\cite{pitchford_qutip_2016}, PICOS 1.1.2~\cite{picos}, and CVXOPT 1.1.9~\cite{cvxopt} running on Python 2.7.12 as provided by the Anaconda distribution.

We compare the performance of the Steane code, the cyclic stabilizer code and the best random code using the above three extrapolated 7-qubit noise models. 
The performance of the codes is shown in \autoref{subfig:innsbruck}. 
First, we see that for a noise model with only single- and two-qubit errors (the convex extrapolation), the cyclic code always outperforms the Steane code. 
This performance is as expected, as the code was tailored to such errors. 
When high-weight errors dominate (as in the product extrapolation), the Steane code outperforms the cyclic code and in fact performs at a level comparable to the best selected random code. 
However, the overall logical noise rates for this channel (at more than $0.5$) are unrealistically large and suggest that this model is a poor extrapolation of the two-qubit data. 
For the intermediate convex-product extrapolated noise model, the Steane code and the cyclic code are very comparable in performance. 
Although the performance of the cyclic code from \autoref{eq:Cyclic} varies greatly for these channels, we nonetheless clearly see the utility of tailoring the code for the specific observed error model. 
The full details of our comparison are provided in the supplementary material as a MATLAB R2016b implementation \cite{supplemental}.

%----------------------------------------------------------------------------------------------------------------------

\section{Discussion} 

We have shown the potential benefits for near-term experimental demonstrations of a quantum memory that are available by tailoring the choice of quantum code to the relevant error model.  More generally, new tailored codes can be found by searching over random codes, subject to desirable constraints such as having low-weight stabilizers.
In particular, one can expect an immediate gain in the performance of these quantum memories with no additional cost in experimental complexity if the search constraints are designed appropriately. 
For example, the 7-qubit quantum error correction demonstration in the ion trap experiment of \citet{Nigg_2014} is predicted to reduce the logical error rate in half with the 7-qubit cyclic code presented here, compared with the Steane code, provided that the frequency of high-weight errors is sufficiently low. 
(Note that, if the correlated error rate is significant, then a natural tailored code may be that studied by~\citet{li_fault_2017}.)
The particular advantage we have demonstrated can be understood as the result of balancing the way that the syndromes respond to the dominant source of errors. 
This is actually a very general approach, and it is possible that simple gains could be made simple by changing stabilizers in known codes using local gates, or by using equivalent codes that have been tailored to the noise bias, as in the work of~\citet{Tuckett_2017}.
This analysis also motivates further research on methods to determine the relevant error model for experiments, in particular in identifying correlated multi-qubit error rates. Another class of experimentally-relevant errors are coherent (or unitary) errors, and it would be worthwhile to determine if codes can be tailored for particular coherent error models. Finally,
 another avenue for further work using this approach would be to incorporate aspects of fault-tolerance such as the presence of specific transversal gates. Such gates may be included as a constraint in the numerical search over random codes.

\begin{acknowledgments}
	The authors are grateful to David Poulin for helpful discussions and Rainer Blatt, Thomas Monz, and Daniel Nigg for sharing the data from the ion trap experiment. 
	This work is supported by the Australian Research Council (ARC) via the Centre of Excellence in Engineered Quantum Systems (EQuS) project number CE110001013 and by the US Army Research Office grant numbers W911NF-14-1-0098 and W911NF-14-1-0103. 
	STF was supported by the ARC via Future Fellowship FT130101744. 
\end{acknowledgments}

\bibliography{TailoredCodes}

\end{document}